\documentclass{aastex}
\usepackage{natbib}

\shorttitle{New very LIS's for cosmic rays}
\shortauthors{Bisschoff, Potgieter \& Aslam}

\begin{document}

\title{{\bf{New very local interstellar spectra for electrons, positrons, protons and light cosmic ray nuclei}}}

\author[0000-0001-7623-9489]{D. Bisschoff}
\affil{Centre for Space Research, North-West University, 2520 Potchefstroom, South Africa}
\email{driaan.b@gmail.com}

\author[0000-0003-0793-7333]{M.S. Potgieter}
\affil{Centre for Space Research, North-West University, 2520 Potchefstroom, South Africa}

\author[0000-0001-9521-3874]{O.P.M.  Aslam}
\affil{Centre for Space Research, North-West University, 2520 Potchefstroom, South Africa}

\begin{abstract}

The local interstellar spectra (LIS's) for galactic cosmic rays (CR's) cannot be directly observed at the Earth below certain energies, because of solar modulation in the heliosphere. With Voyager 1 crossing the heliopause in 2012, in situ experimental LIS data below 100 MeV/nuc can now constrain computed galactic CR spectra. Using galactic propagation models, galactic electron, proton and light nuclei spectra can be computed, now more reliably as very LIS’s. Using the Voyager 1 observations made beyond the heliopause, and the observations made by the PAMELA experiment in Earth orbit for the 2009 solar minimum, as experimental constraints, we simultaneously reproduced the CR electron, proton, Helium and Carbon observations by implementing the GALPROP code. Below about 30 GeV/nuc solar modulation has a significant effect and a comprehensive three-dimensional (3D) numerical modulation model is used to compare the computed spectra with the observed PAMELA spectra possible at these energies. Subsequently the computed LIS's can be compared over as wide a range of energies as possible. The simultaneous calculation CR spectra with a single propagation model allows the LIS's for positrons, Boron and Oxygen to also be inferred. This implementation of the most comprehensive galactic propagation model (GALPROP), alongside a sophisticated solar modulation model to compute CR spectra for comparison with both Voyager 1 and PAMELA observations over a wide energy range, allows us to present new self-consistent very LIS's (and expressions) for electrons, positrons, protons, Helium, Carbon, Boron and Oxygen for the energy range of 3 MeV/nuc to 100 GeV/nuc.

\end{abstract}


\section{Introduction}

Because of solar modulation, the local interstellar spectra (LIS's) for cosmic rays (CR's) are still not fully determined over all energy ranges. A breakthrough came when Voyager 1 (V1) made CR measurements beyond the heliopause in August 2012 for the first time \citep{Voyager1,Gurnett2013,WebberMcDonald2013,Cummings2016}. With its unique positioning, the V1 observations have become central to recent lower energy LIS studies, allowing for galactic electrons, protons and light nuclei (specifically Helium, Carbon and Oxygen) LIS's to be determined with vastly more confidence down to a few MeV/nuc (about 3 MeV/nuc for protons and nuclei, and about 2.7 MeV for electrons \citep{Cummings2016}).

For much higher energies very precise CR spectra are observed at the Earth by experiments such as the satellite-borne PAMELA instrument \citep{Adriani2011,PamelaReview2014,Boezio2015,Boezio2017}. PAMELA made highly valuable observations during the 2006-2009 solar minimum, with specifically the end of 2009 showing unusually high CR intensities
\citep[see e.g.][]{Heber2009,Mewaldt2010,Aslam2012,Lave2013,PotgieterModel2014}, which we consider ideal for estimating the CR LIS's at these high energies. The observations made by PAMELA are over an energy range of about 80 MeV/nuc to 50 GeV/nuc for protons and 80 MeV to 40 GeV for electrons. For energies above about 30 GeV/nuc, the CR spectra can be considered directly observed LIS's as here modulation is largely negligible, while below this energy the effects of modulation become increasingly pronounced and cannot be neglected.
The PAMELA observations of interest to this study are those made at the end of 2009 for electrons \citep{PAMelecdata2009}, positrons \citep{RiccardoPHD} and protons \citep{PAMprotdata2009}, as well as the those presented as averaged over the 2006-2009 period for positrons \citep{PAMposidata}, Helium \citep{PAMhelidata} and Carbon \citep{PAMcarbdata}.

A comprehensive galactic propagation model, such as the GALPROP code  \citep{StrongMoskalenko1998b,StrongReview2007,Porter2017}, can be used to compute galactic spectra for a wide set of CR species.
GALPROP has been used for various aspects of CR spectrum studies, including the study of CR electrons \citep{Strong2011}, CR anti-particles \citep{Moskalenko2002}, galactic turbulence and CR diffusion \citep{Ptuskin2006}, and the galactic broadband luminosity spectrum \citep{Strong2010}.
Other propagation codes have also been used to similarly compute galactic spectra, assumed to be LIS's. A simple leaky box model by \citet{Webber2009} has been used to compute electron \citep{Webber2015elec,WebberVilla2017} and light nuclei \citep{Webber2015prot} LIS's separately. Extensive studies on CR isotopes were done by \citet{Lave2013} using this type of model.
Other more advanced models focused on specific features such as the effects of discrete CR source distributions by \citet{Busching2005} and the effect of the galactic spiral arm structure investigated by \citet{Busching2008} and \citet{Kopp}. Improving the implementation of physical processes has also been investigated, such as incorporating fully anisotropic diffusion by \citet{Effenberger}.
While codes, such as presented by \citet{Kissmann}, are being developed to improve on the numerical capabilities of galactic propagation models, GALPROP remains the most sophisticated and comprehensive galactic propagation model. The history of GALPROP, together with the accessibility offered by GALPROP's dedicated WebRun service \citep{Webrun}, lead the authors to believe that it is the best suited model for a LIS study with a wide scope of CR particles such as this.

After implementing the propagation model, the numerically computed LIS's can then be matched against available observations in order to verify the spectra. For the V1 and higher energy PAMELA observations the comparison can be done directly, but for Earth-based observations below about 30 GeV/nuc the effects of solar modulation need to be taken into account. In order to do this and compute the CR spectra inside the heliosphere at the Earth's position, the sophisticated 3D modulation code presented by \citet{PotgieterVos2017} is used. This model has previously been used to compute electron \citep{PotgieterVoselectable} and proton \citep{VosPotgieterprottable} spectra with older LIS's.
Modulation models require a spectrum to be specified at the chosen modulation boundary, the heliopause, as initial input condition \citep[see the reviews by][]{PotgieterReview2013,Potgieter2017}. The propagation models compute a galactic spectrum for a given CR species, which is strictly speaking not an exact representation of the CR intensities in the local vicinity of the heliosphere. When not including local CR sources and local galactic structures, a computed galactic spectra may not be the same as a very local interstellar spectrum (the required heliopause spectrum). However, for this study the computed galactic spectra will be taken as a heliopause spectra for input into the modulation model, and the GALPROP parameters presented here take this into account. For further discussion on galactic spectra, LIS's and heliopause spectra, from a solar modulation point of view, see \citet{Potgieter2014Braz}.

This study is considered to be the first to concurrently implement these comprehensive models for CR propagation in the Galaxy and heliosphere simultaneously for electrons, positrons, protons, Helium, Carbon, Boron and Oxygen. Using the relevant observations made by V1 and PAMELA, LIS's can be computed with more confidence than before in the energy range of 3 MeV/nuc to 100 GeV/nuc, even for the energies that are not directly observed.

\section{The numerical propagation model and assumptions}

The GALPROP code computes the propagation of relativistic charged particles through the Galaxy, by describing this propagation as a diffusive process. The GALPROP model implements the following equation for the propagation of a particular particle species:
\begin{equation}
\label{galacticprop}
\frac{\partial \psi}{\partial t} = S(\mathbf{r},p)+\nabla\cdot(K\nabla\psi-\mathbf{V}\psi) + \frac{\partial}{\partial p}\left[p^2 K_{p}\frac{\partial}{\partial p}\frac{1}{p^2}\psi + \frac{p}{3}(\nabla\cdot\mathbf{V})\psi - \dot{p}\psi\right] - \frac{\psi}{\tau},
\end{equation}
where $\psi = \psi(\mathbf{r},p,t)$ is the CR density per unit of total particle momentum $p$ at position  $\mathbf{r}$ \citep{StrongReview2007}. The source term $S(\mathbf{r},p)$ includes contributions from primary sources, as well as spallation and decay. Spatial diffusion is represented by the coefficient $K$. The spatially dependent convection velocity is represented by $\mathbf{V}$ and determined by the gradient in the galactic wind d$V$/d$z$. The $\nabla\cdot\mathbf{V}$ term represents the adiabatic momentum gain or loss in the non-uniform flow of gas with a frozen-in magnetic field whose inhomogeneities scatter the CRs. Reacceleration is described as diffusion in momentum space and determined by the coefficient $K_{p}$, while $\dot{p}$
is the momentum loss rate and comprises all forms of energy losses, mostly by synchrotron radiation for electrons or ionization loss for protons and heavier nuclei. The catastrophic particle losses are represented by the timescale $\tau$, which includes the timescale for fragmentation ($\tau_f$), which depends on the total spallation cross-section, and the timescale for radioactive decay ($\tau_r$).

In GALPROP the spatial diffusion coefficient $K$ is assumed to be independent of radius $r$ and height $z$, and is taken as being proportional to a power-law in rigidity $P$:
\begin{equation}
\label{galacticdiff}
K=\beta K_{0}(P /P_{0})^{\delta},
\end{equation}
where $\delta=\delta_{1}$ for rigidity $P < P_{0}$ (the reference rigidity), while $\delta=\delta_{2}$ for $P > P_{0}$. Here  $\beta = v/c$ is the speed of particles $v$ at a given rigidity relative to the speed of light $c$. The magnitude of the diffusion coefficient $K_0$ is in effect a scaling factor for diffusion, generally with units of 10$^{28}$\,cm$^{2}$\,s$^{-1}$. When considering reacceleration, the momentum-space diffusion coefficient $K_{p}$ is estimated as related to $K$ so that $K_{p} K\propto p^2 V_A ^2$,
with $V_A$ the Alfve\'n wave speed set to 36\,km\,s$^{-1}$.

The CR sources are assumed to be concentrated near the galactic disk and have a radial distribution similar to that of supernova remnants, with the distribution assumed to be the same for all CR primaries. The primary contribution to the sources requires an injection spectrum and relative isotopic compositions to be specified. The injection spectrum for nuclei, as input to the source term, is assumed to be a power-law in rigidity so that:
\begin{equation}
\label{galacticsource}
S(P) \propto (P /P_{\alpha 0})^{\alpha},
\end{equation}
for the injected particle density and usually contains a break in the power-law with index $\alpha = \alpha _1$ below the source reference rigidity $P_{\alpha 0}$ and $\alpha = \alpha _2$ above. Values for $\alpha _1$ and $\alpha _2$ are positive and non-zero, thus giving a rigidity dependent injection spectrum. For isotopes considered wholly secondary, no input spectrum is given and they are set to zero at the sources. For the source abundance values as used in GALPROP, this study keeps the values unchanged from what is used by \citet{Ptuskin2006}. The source spectrum is the same for all CR nuclei, but differs for electrons. The injection spectrum for electrons is input similarly to that of the CR nuclei:
\begin{equation}
\label{galacticelecsource}
S(P) \propto (P /P_{\alpha e0})^{\alpha _e},
\end{equation}
where $\alpha _e = \alpha _{e1}$ below the source reference rigidity $P_{\alpha e0}$ and $\alpha _e = \alpha _{e2}$ above \citep{StrongReview2007}.

The Galaxy is described as a cylindrical volume for CR propagation studies. This includes a galactic halo, in which CRs have a finite chance to return to the galactic disk. Assuming symmetry in azimuth leads to a two spatial dimensional (2D) model that depends only on galactocentric radius and height, additionally, neglecting time dependence leads to a steady-state model. When implemented in the GALPROP code this gives a 2D model with radius $r$, the halo height $z$ above the galactic plane and symmetry in the angular dimension in galactocentric-cylindrical coordinates. The propagation region is bounded by $r = R$ and $z = \pm H$, beyond which free escape is assumed. For this study the halo height is fixed at 4 kpc, because varying its size can simply be counteracted by directly varying the diffusion coefficient. While the GALPROP code has been designed for the propagation of CRs on either a 2D or 3D spatial grid \citep{StrongMoskalenko2001}, only the 2D model is considered for this study, that is, two spatial dimensions and momentum, giving the basic coordinates $(r,z,p)$ for the rotationally symmetric cylindrical grid. Symmetry is assumed above and below the galactic plane in order to save on the computational requirements of the code. The GALPROP code solves the propagation equation, for each of the CR species that are taken into account, using a Crank-Nicholson implicit second-order scheme. The processes are described by differential operators in the propagation equation and these operators are implemented as finite differences for each dimension $(r,z,p)$ in the numerical scheme.
For extensive details on solving the propagation equation, the numerical scheme and differential operators see  \citet{StrongMoskalenko1998b,StrongReview2007,Supplement}. The computational runs for this study are done via the GALPROP WebRun service, offers the benefits of running the most recent version of GALPROP, with error detection, powerful computing power and user support. The service can be accessed at: \url{http://galprop.stanford.edu/webrun}. The details on the implementation of the WebRun service, the updated features of the code and the computer cluster specifications are presented by \citet{Webrun}.

\section{The numerical transport model for solar modulation}

In order to compute the spectra of CRs as they arrive at the Earth, after their transport through the heliosphere, a numerical modulation model is used. This full 3D model, as described by \citet{PotgieterVos2017}, can compute modulated differential intensities throughout the heliosphere by implementing the CR transport equation first derived by \citet{Parker1965}. This model takes into account the major modulation mechanisms of convection and adiabatic energy losses due to the expanding solar wind, particle diffusion and drifts due to the heliospheric magnetic field. This steady-state model also includes a wavy current sheet and a heliosheath, but does not consider shock acceleration at the termination shock. For the purposes of this work, we restrict the modulation, with a few exceptions, to the PAMELA observations made during the solar minimum of 2009, which was an A$< 0$ solar magnetic field epoch.

The CR transport equation by \citet{Parker1965} can be written in terms of rigidity ($P$) as :
\begin{equation}
\frac{\partial f}{\partial t} = - {\mathbf V} _{sw} \cdot \nabla f - \langle {\mathbf v} _{D} \rangle \cdot \nabla f + \nabla \cdot ({\mathbf K} _{s} \cdot \nabla f) + \frac {1}{3} (\nabla \cdot \mathbf{V}_{sw}) \frac{\partial f}{\partial  \ln  P} \, ,
\label{Eq1}
\end{equation}
where $f(\mathbf{r}, P, t)$ is the CR distribution function at time $t$ and at vector position $\mathbf{r}$.
For the left side of the equation $\frac{\partial f}{\partial t} = 0$ as we consider only a steady-state solution for the solar minimum conditions where the modulation parameters only gradually change over time. The first term on the right gives the outward convection due to the solar wind. The second term represents the averaged particle drift velocity $\langle \mathbf{v}_{D} \rangle$.
\begin{equation}
\langle \mathbf{v}_{D} \rangle = \nabla \times K_{A} \frac{\mathbf{B}}{B},
\end{equation}
where $K_{A}$ is the generalized drift coefficient, and $\mathbf{B}$ is the heliospheric magnetic field (HMF) vector with magnitude $B$. The third term describes the spatial diffusion caused by the scattering of CRs, where $\mathbf{K}_{s}$ is the symmetric diffusion tensor, and the last term represents the adiabatic energy change, which depends on the sign of the divergence of $\mathbf{V}_{sw}$.
If ($\nabla \cdot \mathbf{V}_{sw}) > 0$ adiabatic energy losses occur, as is the case in most of the heliosphere, except inside the heliosheath where we assume that $(\nabla \cdot \mathbf{V}_{sw}) = 0$.
For a modified Parker-type HMF, $\mathbf{B}_{m}$ with magnitude $B_{m}$, such as the Smith-Bieber modification \citep{SmithBieber1991}, the drift coefficient $K_{A}$ can be written as:
\begin{equation}
K_{A} = \frac {\beta P} {3B_{m}} f_{D} = K_{A0}  \frac {\beta P} {3B_{m}} \frac {(P/P_{A0})^{2}}{1+(P/P_{A0})^{2}}.
\end{equation}
The dimensionless constant $K_{A0}$ ranges from 0.0 to 1.0, where if $K_{A0}$ = 1.0 is called 100\% drift or full weak scattering. In this study $K_{A0}$ is kept at 0.90, effectively setting particle drift to a 90\% level. For detailed discussions of this process, see also for example, \citet{Ngobeni2015}, \citet{Nndanganeni2016} and \citet{Raath2016}.

The symmetric diffusion tensor $\mathbf{K} _{s}$ is comprised of three diffusion coefficients, $K_{\parallel}$, $K_{\perp r}$ and $K_{\perp \theta}$.
The expression for the diffusion coefficient parallel to the average background HMF is given by:
\begin{equation}
K_{\parallel} = (K_{\parallel})_{0}\, \beta \left(\frac {B_{0}}{B}\right) \left(\frac {P}{P_{0}}\right)^{c_{1}} \left( \frac {\left(\frac {P}{P_{0}}\right)^{c_{3}} +
 \left(\frac{P_{k}}{P_{0}}\right)^{c_{3}}}{ 1+ \left(\frac {P_{k}}{P_{0}}\right)^{c_{3}}} \right)^{\frac {c_{2 \parallel} - c_{1}}{c_{3}}},
\label{EqK}
\end{equation}
with $(K_{\parallel})_{0}$ a scaling constant in units of $10^{22}$ cm$^{2}$s$^{-1}$, $P_{0}$ = 1\,GV and $B_{0}$ = 1\,nT. The power indices $c_{1}$ and $c_{2 \parallel}$ respectively determine the slope of the rigidity dependence of $K_{\parallel}$ above and below the rigidity $P_{k}$, while $c_{3}$ determines the smoothness of the transition.
Perpendicular diffusion in the radial direction ($K_{\perp r}$) is assumed to scale spatially similar to Equation \ref{EqK}, but with a different rigidity dependence at higher rigidities so that:
\begin{equation}
K_{\perp r} = 0.02\,(K_{\parallel})_{0}\, \beta \left(\frac {B_{0}}{B}\right) \left(\frac {P}{P_{0}}\right)^{c_{1}} \left( \frac {\left(\frac {P}{P_{0}}\right)^{c_{3}} +
 \left(\frac{P_{k}}{P_{0}}\right)^{c_{3}}}{ 1+ \left(\frac {P_{k}}{P_{0}}\right)^{c_{3}}} \right)^{\frac {c_{2 \perp} - c_{1}}{c_{3}}},
\end{equation}
which is close to the widely used assumption of $K_{\perp r} = 0.02\,K_{\parallel}$. The polar perpendicular diffusion coefficient ($K_{\perp \theta}$) is given by:
\begin{equation}
K_{\perp \theta} = 0.02 K_{\parallel} f_{\perp \theta} = K_{\perp r} f_{\perp \theta}.
\end{equation}
The latitudinal dependence ($f_{\perp \theta}$) in the above equation is given by:
\begin{equation}
f_{\perp \theta} = A^{+} \mp A^{-} \tanh[8(\theta_{A} - 90^{\circ}) \pm \theta_{F}],
\end{equation}
where A$^{\pm} = (d_{\perp \theta} \pm 1)/2 $, $\theta_{F}$ = 35$^{\circ}$, $\theta_{A}$ = $\theta$ for $\theta \leq 90^{\circ}$ and $\theta_{A}$ = 180$^{\circ}$ - $\theta$ for $\theta \geq$  90$^{\circ}$.
With this expression $K_{\perp \theta}$ can be enhanced towards the heliospheric poles by a factor d$_{\perp \theta}$. For examples of modulated spectra computed using this model, see the reviews by \citet{PotgieterReview2013, Potgieter2017}, for a comprehensive discussion of charge-sign dependent modulation in the heliosphere, see \citet{Potgieter2014chargesign}, and for details on the 3D model described in short above,
see also \citet{PotgieterModel2014, PotgieterVoselectable, VosPotgieterprottable} and \citet{Aslam2019}.

\begin{figure}[!t]
  \centering
  \epsscale{0.80}
  \plotone{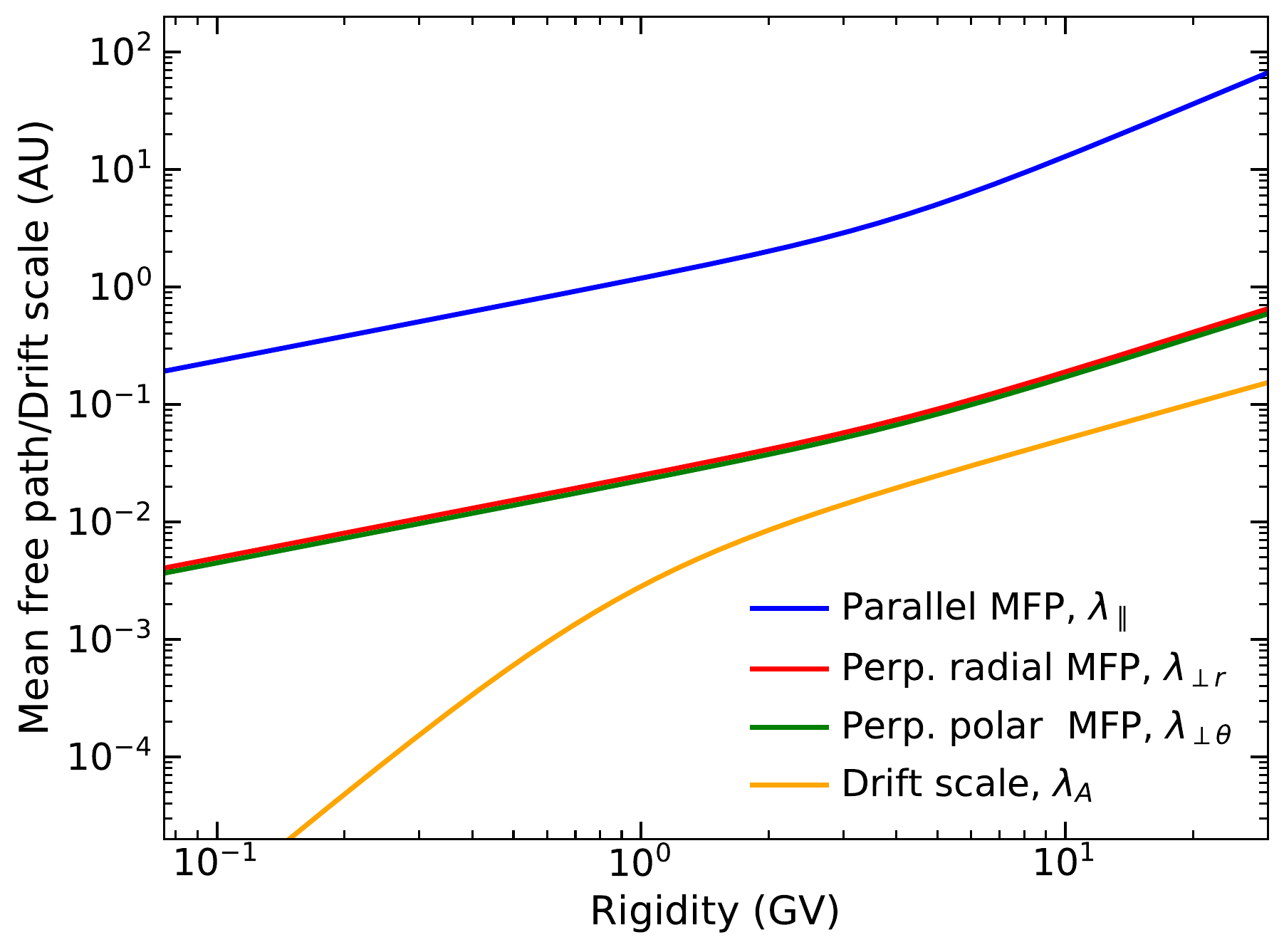}
  \caption{Rigidity dependent MFPs and drift scale used for modulation as adjusted from \citet{VosPotgieterprottable} to modulate the proton and light nuclei LIS's for the 2009b period. Shown is the parallel ($\lambda _{\parallel}$, blue curve), radial perpendicular ($\lambda _{\perp r}$, red curve), polar perpendicular ($\lambda _{\perp \theta}$, green curve) MFPs, and the drift scale (yellow curve). These rigidity dependent values result from the choice of parameters in Table \ref{Table1}.}
  \label{diff_proton}
\end{figure}
\begin{figure}[!b]
  \centering
  \epsscale{0.80}
  \plotone{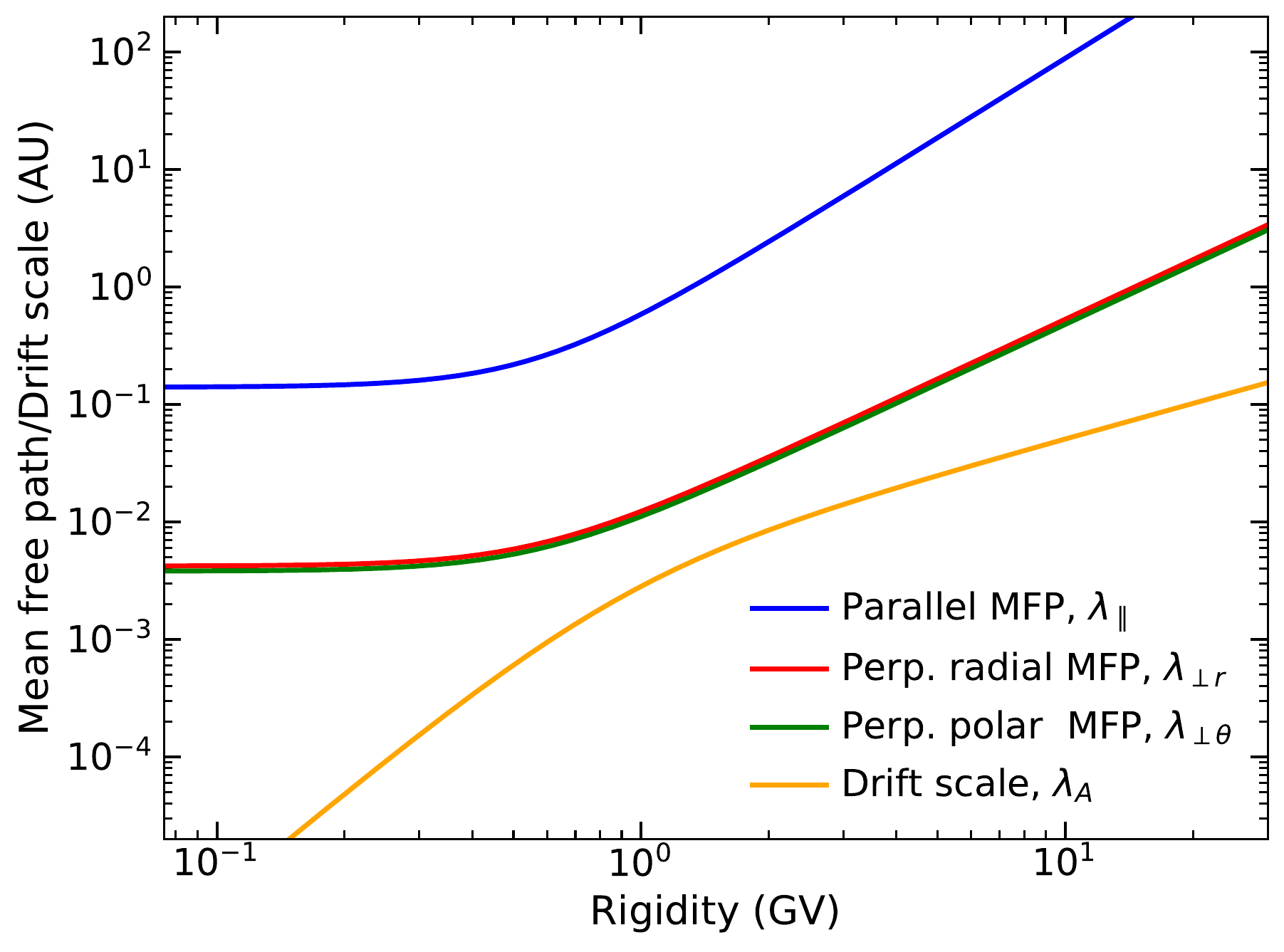}
  \caption{Rigidity dependent MFPs and drift scale used for modulation as adjusted from \citet{PotgieterVoselectable} to modulate the electron and positron LIS's for the 2009b period. Shown is the parallel ($\lambda _{\parallel}$, blue curve), radial perpendicular ($\lambda _{\perp r}$, red curve), polar perpendicular ($\lambda _{\perp \theta}$, green curve) MFPs, and the drift scale ($\lambda _{A}$, yellow curve). These rigidity dependent values result from the choice of parameters in Table \ref{Table1}.}
  \label{diff_electron}
\end{figure}

With this model the proton LIS can be modulated to compute the corresponding spectrum at the Earth. The relevant modulation parameter values required to reproduce the observed PAMELA spectra for the last half of 2009, indicated as 2009b, are summarized in Table \ref {Table1} and are adjusted from those presented by \citet{VosPotgieterprottable}. The diffusion coefficients are related to the corresponding MFPs by, $K$ = $\lambda$($v/3$), where $v$ is the particle speed, giving rigidity dependent MFPs: $\lambda _{\parallel}$ (blue curve), $\lambda _{\perp r}$ (red curve), $\lambda _{\perp \theta}$ (green curve), and the drift scale (yellow curve) as shown in Fig. \ref{diff_proton}. Similarly the light nuclei LIS's can be modulated using the same parameters and accounting for the specific particle masses and charges.
For electrons and positrons the parameters are similarly implemented, as presented by \citet{PotgieterVoselectable}, and have been adjusted to the values listed in Table \ref{Table1}. These values also give the MFPs of Fig. \ref{diff_electron}. The modulated spectra will be shown together with the corresponding LIS's in the rest of the figures in the next section.

\begin{table}[tb]
\caption{Summary of parameters used in the modulation model for the 2009b time period \label{Table1}.}
\centering
\begin{tabular}[b]{ l  c  c}
\hline \hline
Parameters                              & Electrons and & Protons and\\
                                        & Positrons & Light Nuclei\\
\hline
$\lambda _{\parallel 0}$ (AU)          	& 0.593       & 1.185  \\
$K_{A0}$                        				& 0.90        & 0.90        \\
$P_{A0}$ (GV)		                       	& 0.90        & 0.90      \\
$c _1$			                           	& 0.00        & 0.70       \\
$c _{2 \parallel}$			               	& 2.25        & 1.52       \\
$c _{2 \perp}$		                    	& 1.688       & 1.14      \\
$c _3$			                          	& 2.70        & 2.50     \\
$P _k$	(GV)		                       	& 0.57        & 4.00      \\
$d _{\perp \theta}	$		                & 6.00        & 6.00    \\
\hline
\end{tabular}
\end{table}

\begin{table}[tb]
\caption{Summary of parameters used in GALPROP to compute the LIS's of this study \label{Table2}.}
\centering
\begin{tabular}[b]{ l  c  c}
\hline \hline
Parameters                              & Values\\
\hline
$K_0$ (10$^{28}$\,cm$^2$\,s$^{-1}$)   	& 5.1\\
$P_{0}$ (GV)		                       	& 4.0\\
$\delta _1$     		                   	& 0.3\\
$\delta _2$	                            & 0.4\\
$P_{\alpha 0}$ (GV)                     & 9.0\\
$\alpha _1$                             & -1.86\\
$\alpha _2$                             & -2.36\\
$P_{\alpha e 0}$ (GV)                   & 4.0\\
$\alpha _{e1}$                          & -1.9\\
$\alpha _{e2}$                          & -2.7\\
$V_A$ (km\,s$^{-1}$)                     & 30.0\\
d$V/$d$z$ (km\,s$^{-1}$ kpc$^{-1}$)      & 5.0\\
\hline
\end{tabular}
\end{table}

\section{Results}

The LIS's for electrons, positrons and protons were computed first using the values listed by \citet{Ptuskin2006} as reference. The parameters were then systematically adjusted in order to find LIS's that match the mentioned observations. Changes to the electron and nuclei source indices for lower energies were found necessary as to update the GALPROP models to take into account the V1 observations. With diffusion having the largest effect on the LIS's, changes to the diffusion parameters were also required to reproduce the observations well, instead of just a rough reproduction. Initial studies showed that a GALPROP plain diffusion model was sufficient when studying electrons (as presented by \citet{Bisschoff2014}), protons and Helium LIS's separately. Modelling protons, Helium, Carbon and the B/C ratio simultaneously, as presented by \citet{Bisschoff2016}, required the inclusion of reacceleration in galactic space in the plain diffusion model. To further include electrons and positrons, and thus having a model that can simultaneously compute the required set of CR LIS's and match the observations, convection also needed to be taken into account.

\begin{figure}[tb]
  \centering
  \epsscale{0.80}
  \plotone{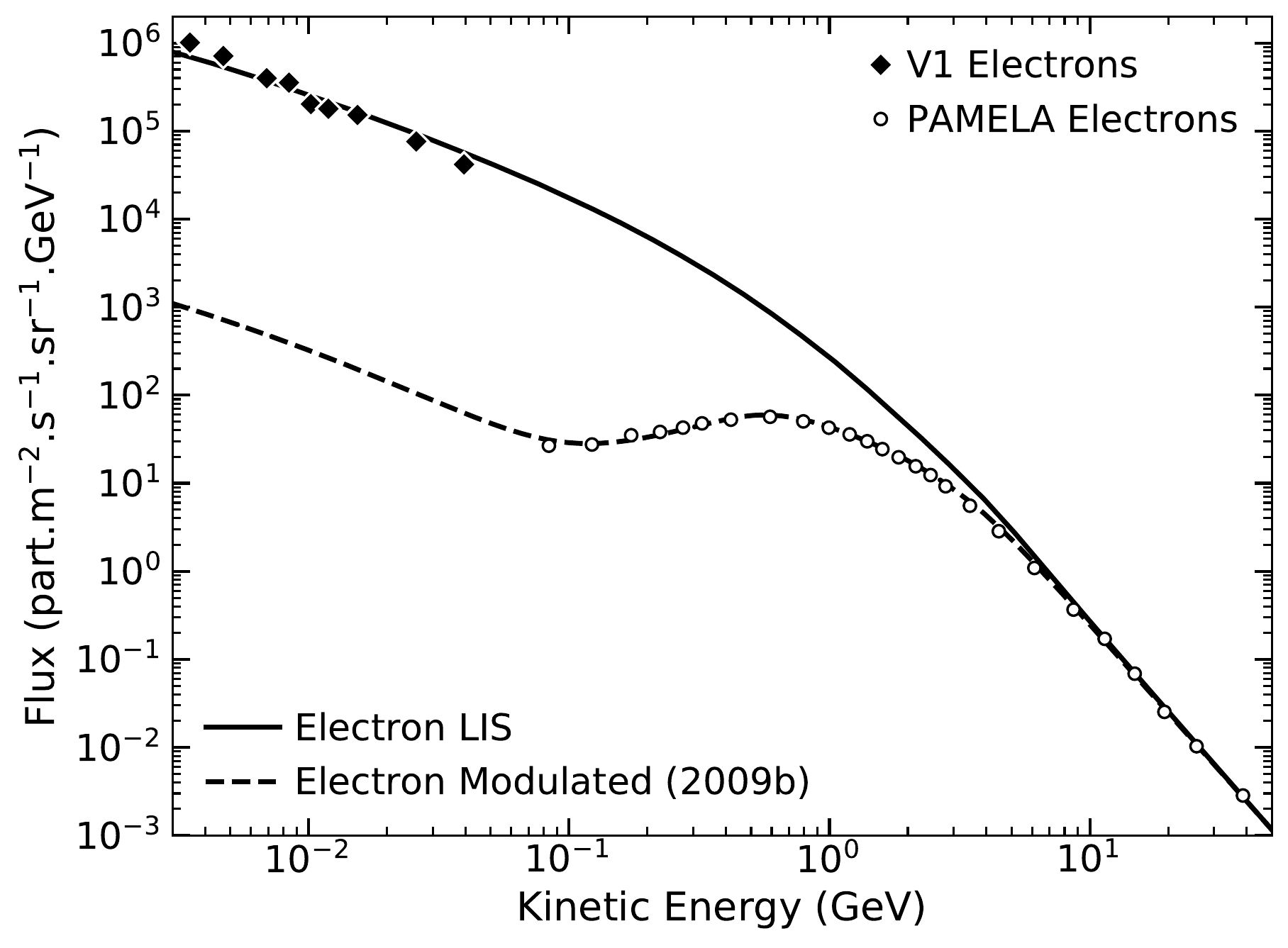}
  \caption{Computed electron LIS (solid black curve) and the corresponding modulated electron spectrum at the Earth (dashed black curve) compared to the V1 electron observations at 122\,AU \citep{Cummings2016} and PAMELA observations at the Earth (1\,AU) for the second half of 2009 \citep{PAMelecdata2009}.}
  \label{FinalElec}
\end{figure}

\begin{figure}[tbp]
  \centering
  \epsscale{0.80}
  \plotone{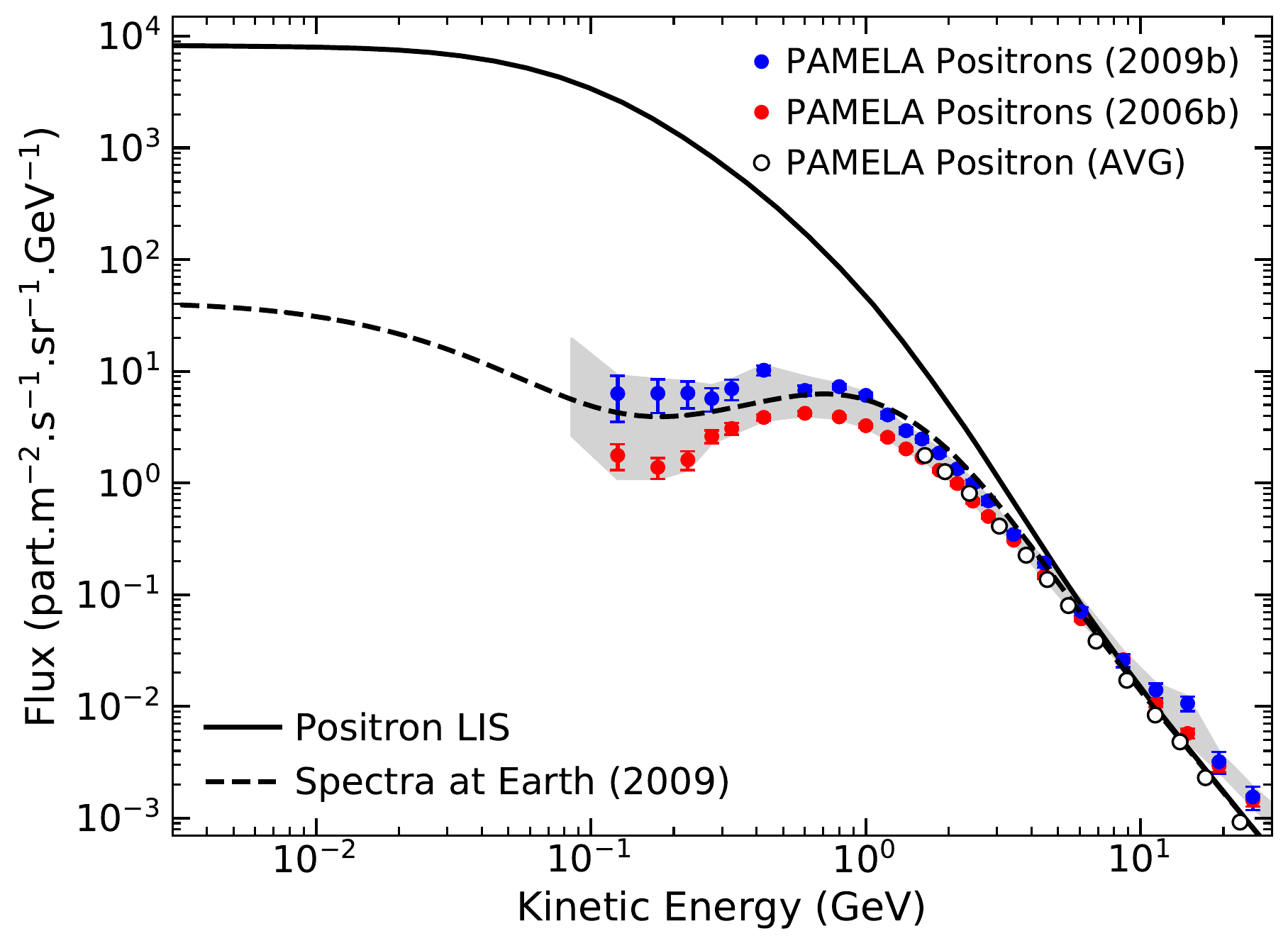}
  \caption{Computed positron LIS (solid black curve) and the computed modulated positron spectrum at the Earth (dashed black curve) compared to the PAMELA positron observations: three year average (open circles) \citep{PAMposidata}, half year averages for July-Dec 2006, indicated as 2006b (red filled circles), and for July-Dec 2009, indicated as 2009b (blue filled circles), together with the maximum variation of the observed positron spectrum within the error margins for Jul 2006 to Dec 2009 (grey band) from \citet{RiccardoPHD}.}
  \label{FinalPosi}
\end{figure}

\begin{figure}[t]
  \centering
  \epsscale{0.80}
  \plotone{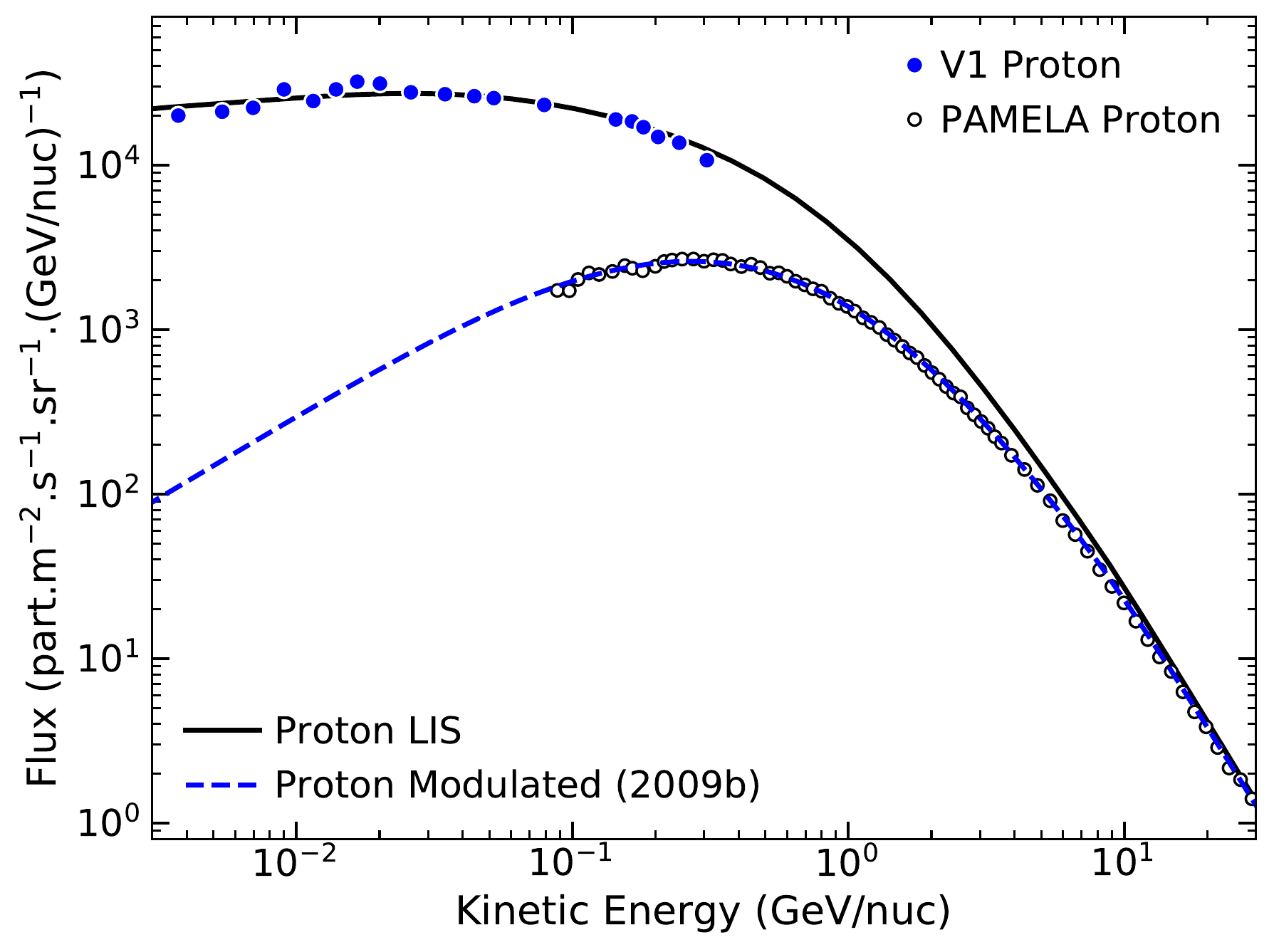}
  \caption{Computed proton LIS (solid black curve) and the computed modulated proton spectrum at the Earth (dashed black curve) compared to the V1 \citep{Voyager1} and PAMELA observations for 2009b \citep{PAMprotdata2009}.}
  \label{FinalProt}
\end{figure}

\begin{figure}[t]
  \centering
  \epsscale{0.80}
  \plotone{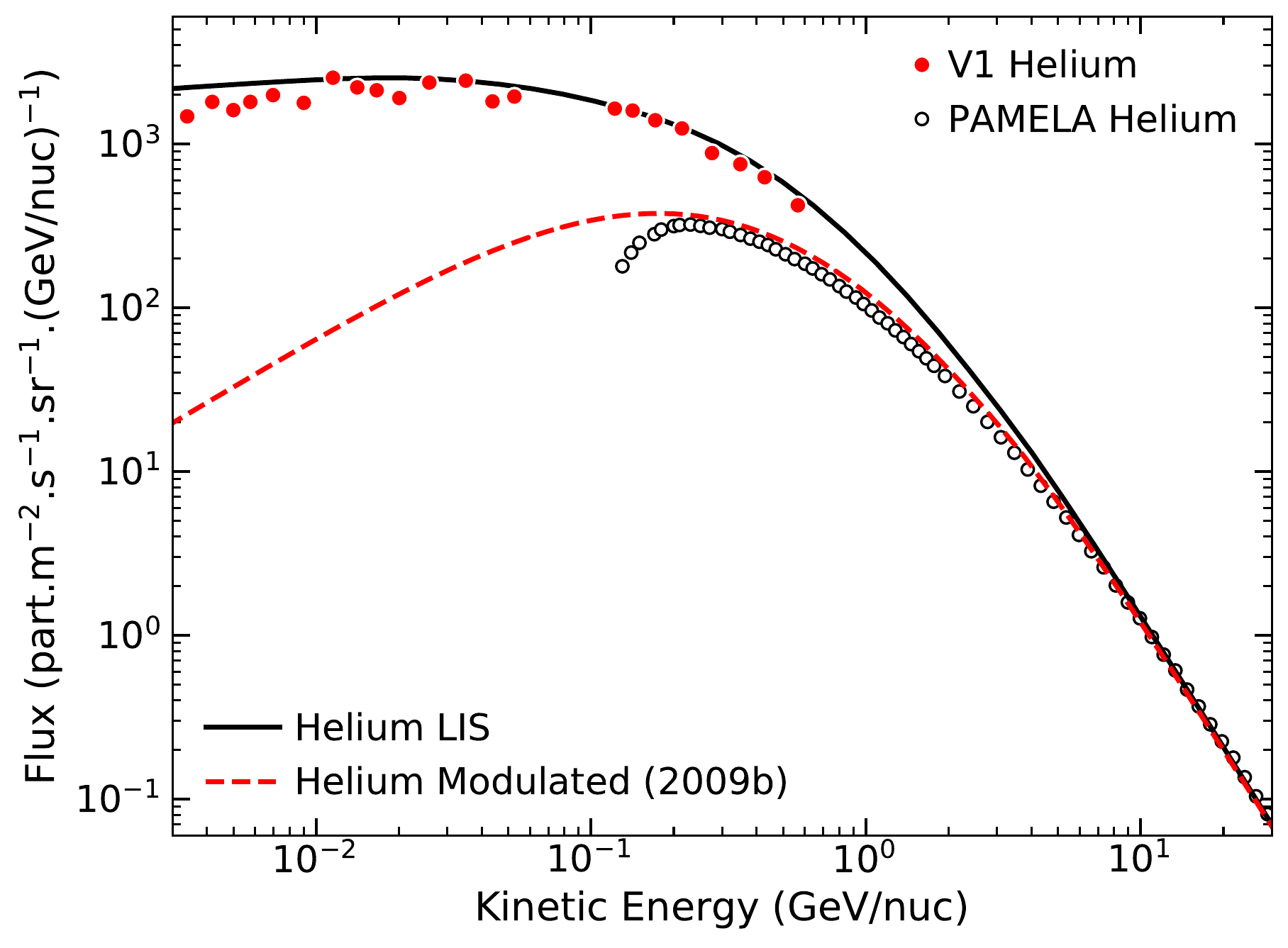}
  \caption{Computed Helium LIS (solid black curve) and the computed modulated Helium spectrum for the period 2009b at the Earth (dashed red curve) compared to the V1 \citep{Voyager1} and PAMELA observations averaged for the Jul 2006 to Dec 2008 period \citep{PAMhelidata}. The observational data for Helium at 1\,AU is currently being updated by the PAMELA group (see \citet{Marcelli2019}).}
  \label{FinalHe}
\end{figure}

\begin{figure}[tbp]
  \centering
  \epsscale{0.80}
  \plotone{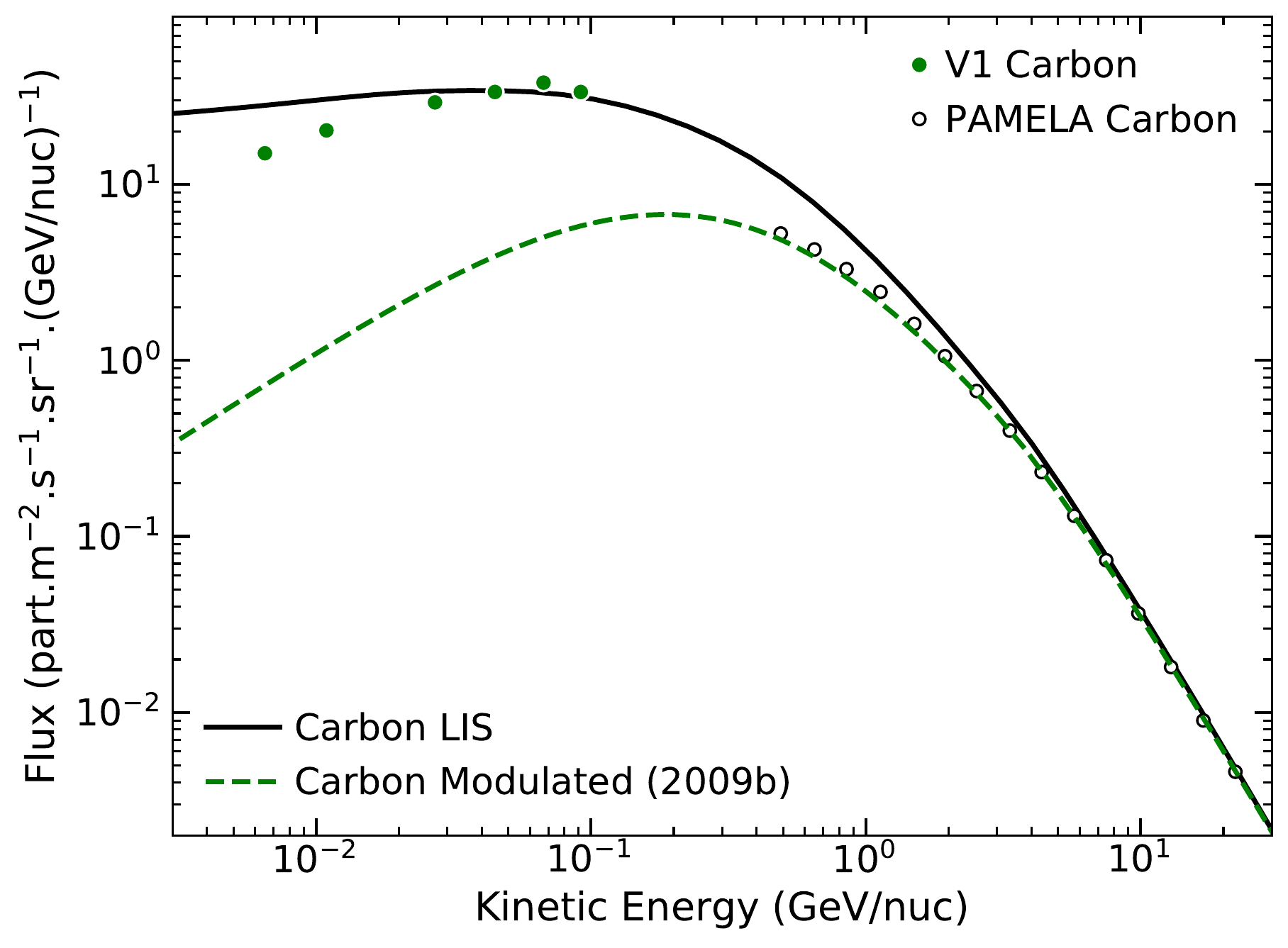}
  \caption{Computed Carbon LIS (solid black curve) and the computed modulated Carbon spectrum at the Earth (dashed green curve) compared to the V1 \citep{Voyager1} and PAMELA observations averaged for the Jul 2006 to Mar 2008 period \citep{PAMcarbdata}.}
  \label{FinalC}
\end{figure}

\begin{figure}[tbp]
  \centering
  \epsscale{0.80}
  \plotone{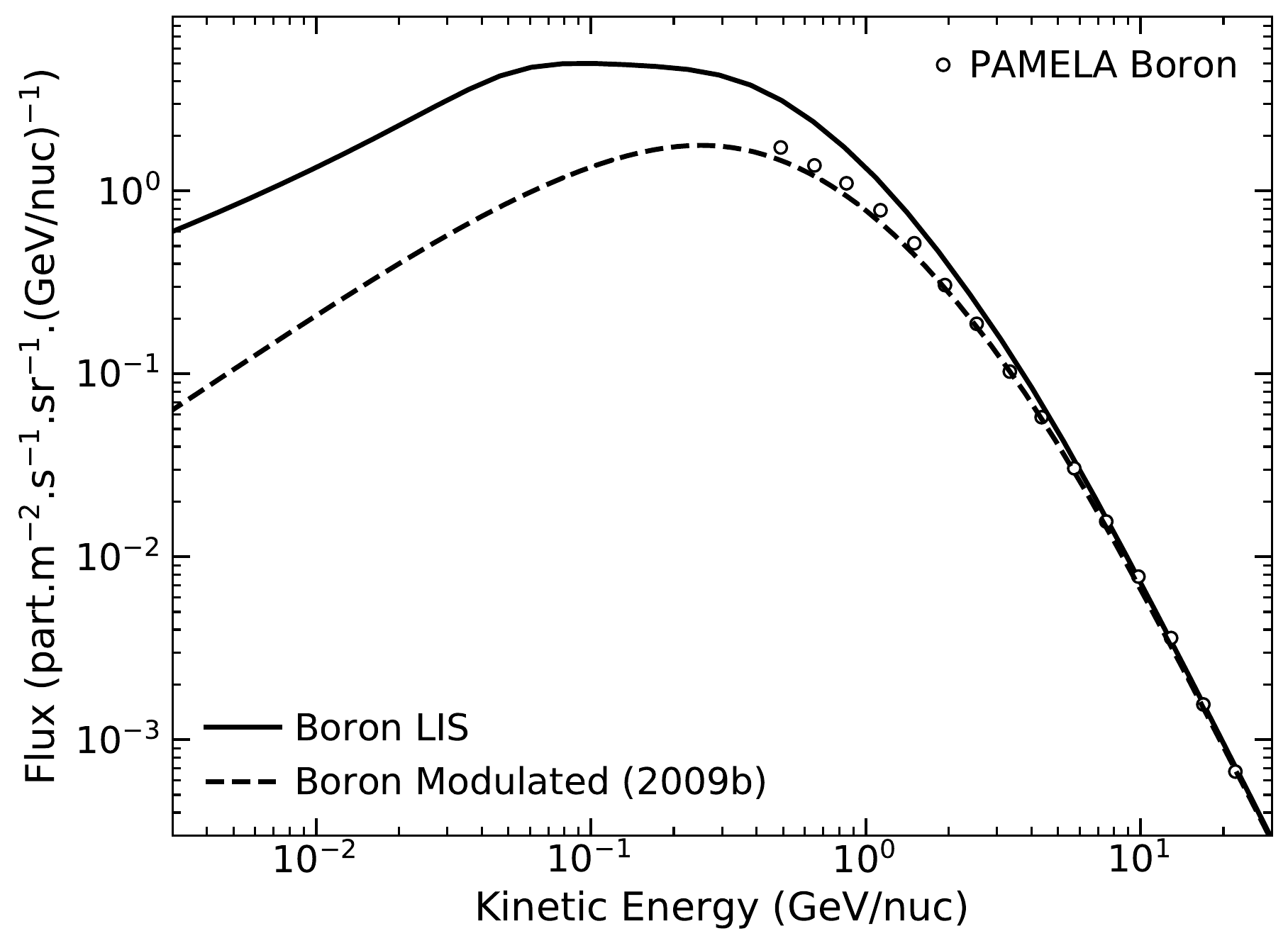}
  \caption{Computed Boron LIS (solid black curve) and the computed modulated Boron spectrum for the period 2009b at the Earth (dashed black curve) compared to the PAMELA observations averaged for the Jul 2006 to Mar 2008 period \citep{PAMcarbdata}.}
  \label{FinalB}
\end{figure}

\begin{figure}[tbp]
  \centering
  \epsscale{0.80}
  \plotone{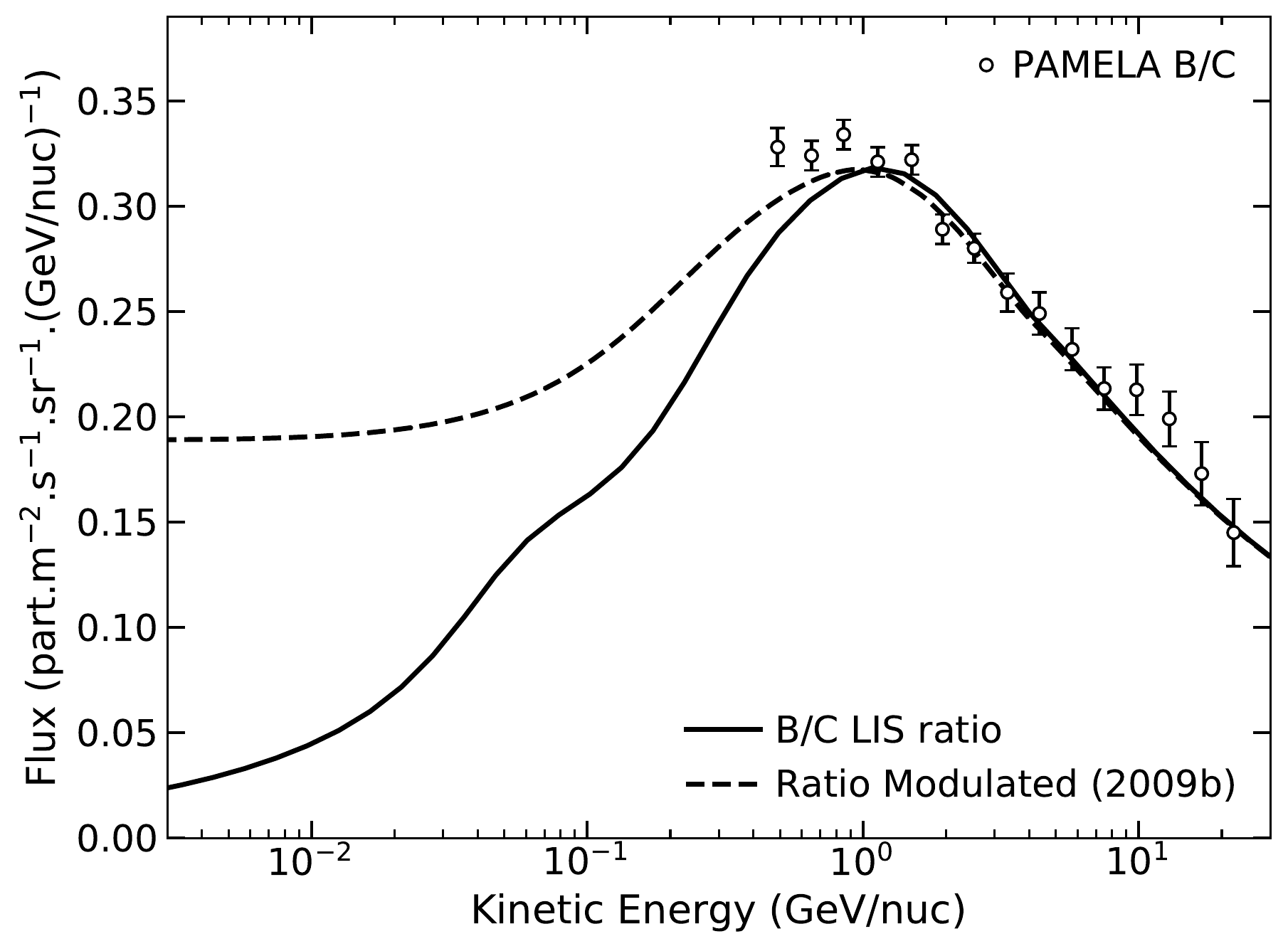}
  \caption{B/C LIS ratio (solid black curve) calculated from the Carbon LIS of Fig. \ref{FinalC} and the Boron LIS of Fig. \ref{FinalB}, compared to the computed B/C ratio for the period 2009b at the Earth (dashed black curve) and the PAMELA B/C observations averaged for the Jul 2006 to Mar 2008 period \citep{PAMcarbdata}.}
  \label{FinalBCrati}
\end{figure}

\begin{figure}[tb]
  \centering
  \epsscale{0.80}
  \plotone{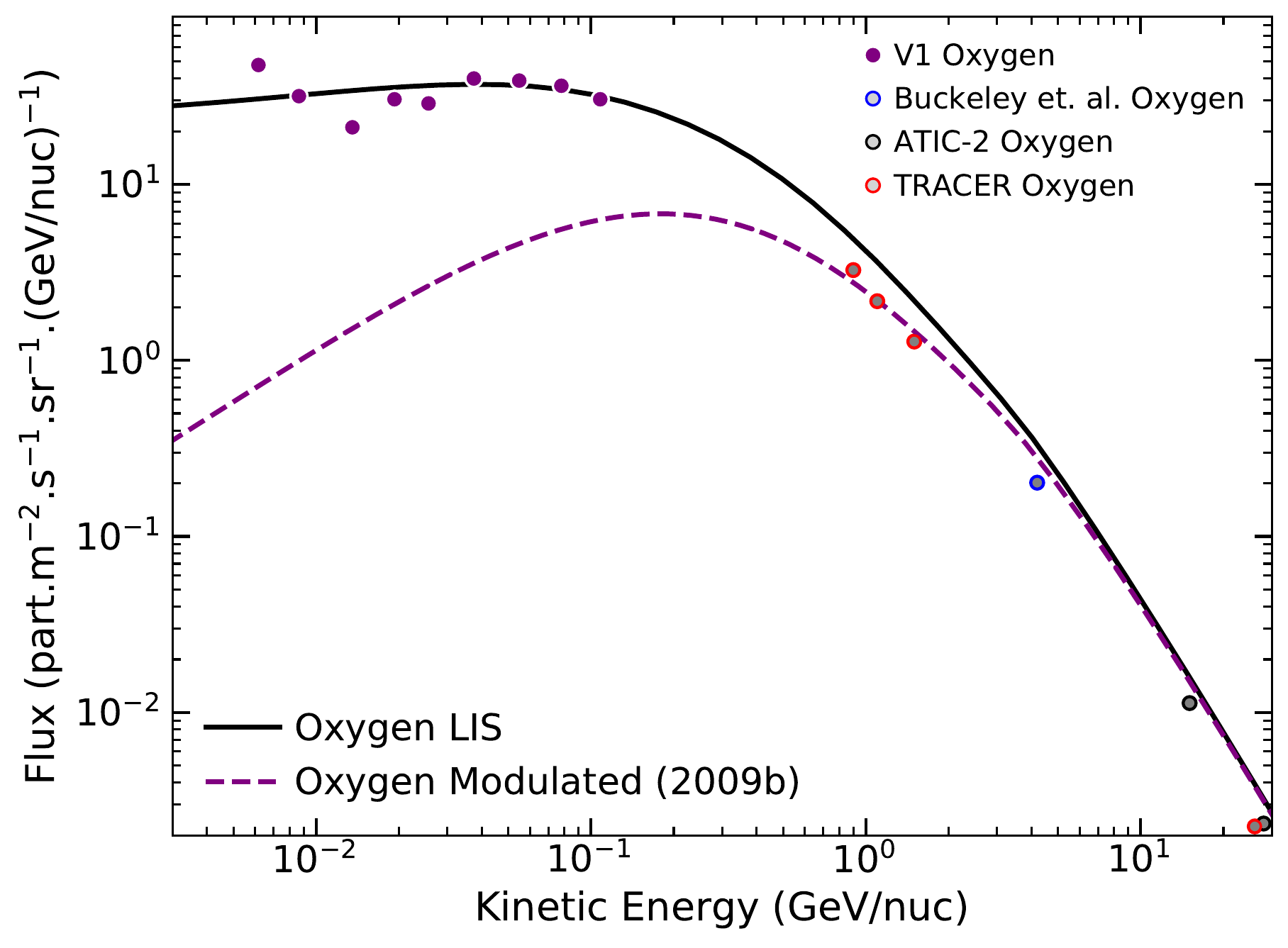}
  \caption{Computed Oxygen LIS (solid black curve) and the computed modulated Oxygen spectrum for the 2009b period at the Earth (dashed purple curve) compared to the V1 observations \citep{Voyager1}. Also shown are the observations made at the Earth by \citet{Buckley} in 1991 (open blue circles), the ATIC-2 experiment \citet{Atic} in 2003 (open black circles) and the TRACER experiment \citet{Tracer} in 2006 (open red circles); all of which were balloon flights.}
  \label{FinalO}
\end{figure}

The LIS's computed with GALPROP for electrons, positrons, protons and the light nuclei are shown in Figs. \ref{FinalElec} to \ref{FinalO}. The parameters used in GALPROP are listed in Table \ref{Table2}, these values were iteratively found and based on the parameter values presented by \citet{Ptuskin2006} and \citet{Strong2010}. Additional changes in GALPROP are made to the source abundances of Helium and Carbon. The PAMELA observations necessitate decreasing the $^{12}$C$_{6}$ abundance by 10\% and increasing the $^4$He$_2$ abundance by 5\%.

The computed electron LIS, shown in Fig. \ref{FinalElec} together with the corresponding V1 observations and the modulated spectrum at the Earth in comparison with PAMELA observations, is approximated over the energy range 4\,MeV to 100\,GeV by the following expression:
\begin{equation}
\label{expr1}
 J_{\rm elec}(E) = 255.0 \, \frac{1}{\beta ^2} \, \left(\frac{E}{E_0}\right) ^{-1} \left(\frac{E/E_0 + 0.63}{1.63} \right) ^{-2.43} + 6.4 \, \left(\frac{E}{E_0}\right) ^{2} \left(\frac{E/E_0 + 15.0}{16.0}\right) ^{-26.0},
\end{equation}
where the CR intensity $J_{\rm elec}(E)$ (given in part.m$^2$\,s$^{-1}$\,sr$^{-1}$\,GeV$^{-1}$) is a function of kinetic energy $E$ (given in GeV), $E_0$ = 1\,GeV and again with $\beta = v/c$.
The modulation parameters are listed in Table \ref{Table1}.

The computed positron LIS and the corresponding modulated spectrum at the Earth, in comparison with PAMELA observations, are shown in Fig. \ref{FinalPosi}. The LIS is approximated by the following expression:
\begin{equation}
\label{expr2}
J_{\rm posi}(E) = 25.0 \, \frac{1}{\beta ^2} \, \left(\frac{E}{E_0}\right) ^{0.1} \left(\frac{(E/E_0)^{1.1} + 0.2^{1.1}}{1 + 0.2^{1.1}} \right) ^{-3.31} + 23.0 \, \left(\frac{E}{E_0}\right) ^{0.5} \left(\frac{E/E_0 + 2.2}{3.2}\right) ^{-9.5}.
\end{equation}

The computed proton LIS and the corresponding modulated spectrum at the Earth, shown in Fig. \ref{FinalProt} in comparison with the corresponding V1 and PAMELA observations, is approximated over the energy range 4\,MeV/nuc to 100\,GeV/nuc by the following expression:
\begin{equation}
\label{expr3}
J_{\rm p}(E) = 2620.0 \, \frac{1}{\beta ^2} \, \left(\frac{E}{E_0}\right) ^{1.1} \left(\frac{(E/E_0)^{0.98} + 0.7^{0.98}}{1 + 0.7^{0.98}} \right) ^{-4.0} + 30.0 \, \left(\frac{E}{E_0}\right) ^{2} \left(\frac{E/E_0 + 8.0}{9.0}\right) ^{-12.0},
\end{equation}
where the CR intensity $J_{\rm p}(E)$ (given in part.m$^2$\,s$^{-1}$\,sr$^{-1}$\,(GeV/nuc)$^{-1}$) is a function of kinetic energy $E$ (given in GeV/nuc), $E_0$ = 1\,GeV/nuc and with $\beta = v/c$.

Similarly, the computed LIS's for Helium, Carbon, Boron and Oxygen are approximated by the following expressions. The Helium LIS, shown in Fig. \ref{FinalHe} together with the corresponding modulated spectrum at the Earth and relevant observations, is approximated by:
\begin{equation}
\label{expr4}
J_{\rm He}(E) = 163.4 \, \frac{1}{\beta ^2} \, \left(\frac{E}{E_0}\right) ^{1.1} \left(\frac{(E/E_0)^{0.97} + 0.58^{0.97}}{1 + 0.58^{0.97}} \right) ^{-4.0}.
\end{equation}
The Carbon LIS, shown in Fig. \ref{FinalC} together with the corresponding modulated spectrum at the Earth and relevant observations, is approximated by:
\begin{equation}
\label{expr5}
J_{\rm C}(E) = 3.3 \, \frac{1}{\beta ^2} \, \left(\frac{E}{E_0}\right) ^{1.22} \left(\frac{(E/E_0)^{0.9} + 0.63^{0.9}}{1 + 0.63^{0.9}} \right) ^{-4.43}.
\end{equation}
The Boron LIS, shown in Fig. \ref{FinalB} together with the corresponding modulated spectrum at the Earth and relevant observations, is approximated by:
\begin{equation}
\label{expr6}
J_{\rm B}(E) =  \frac{1}{\beta ^2} \, \left(\frac{E}{E_0}\right) ^{1.7} \left(\frac{E/E_0 + 0.685}{1.685} \right) ^{-4.8} + (3.0\times\!10^{-4}) \, \left(\frac{E}{E_0}\right) ^{3} \left(\frac{E/E_0 + 0.204}{1.204}\right)^{-11.0}.
\end{equation}
These two LIS's are used to calculate the B/C ratio for the LIS's as shown in Fig. \ref{FinalBCrati}, in comparison with the ratio at the Earth calculated from the modulated spectra. Lastly the Oxygen LIS, shown in Fig. \ref{FinalO} together with the corresponding modulated spectrum at the Earth and relevant observations, is approximated by:
\begin{equation}
\label{expr7}
J_{\rm O}(E) = 3.3 \, \frac{1}{\beta ^2} \, \left(\frac{E}{E_0}\right) ^{1.23} \left(\frac{(E/E_0)^{0.86} + 0.62^{0.86}}{1 + 0.62^{0.86}} \right) ^{-4.62}.
\end{equation}
A more detailed description of these LIS's can be found in the Ph.D. by \citet{BisschoffPHD}.

\section{Discussion and Conclusions}

\begin{figure}[tb]
  \centering
  \epsscale{0.80}
  \plotone{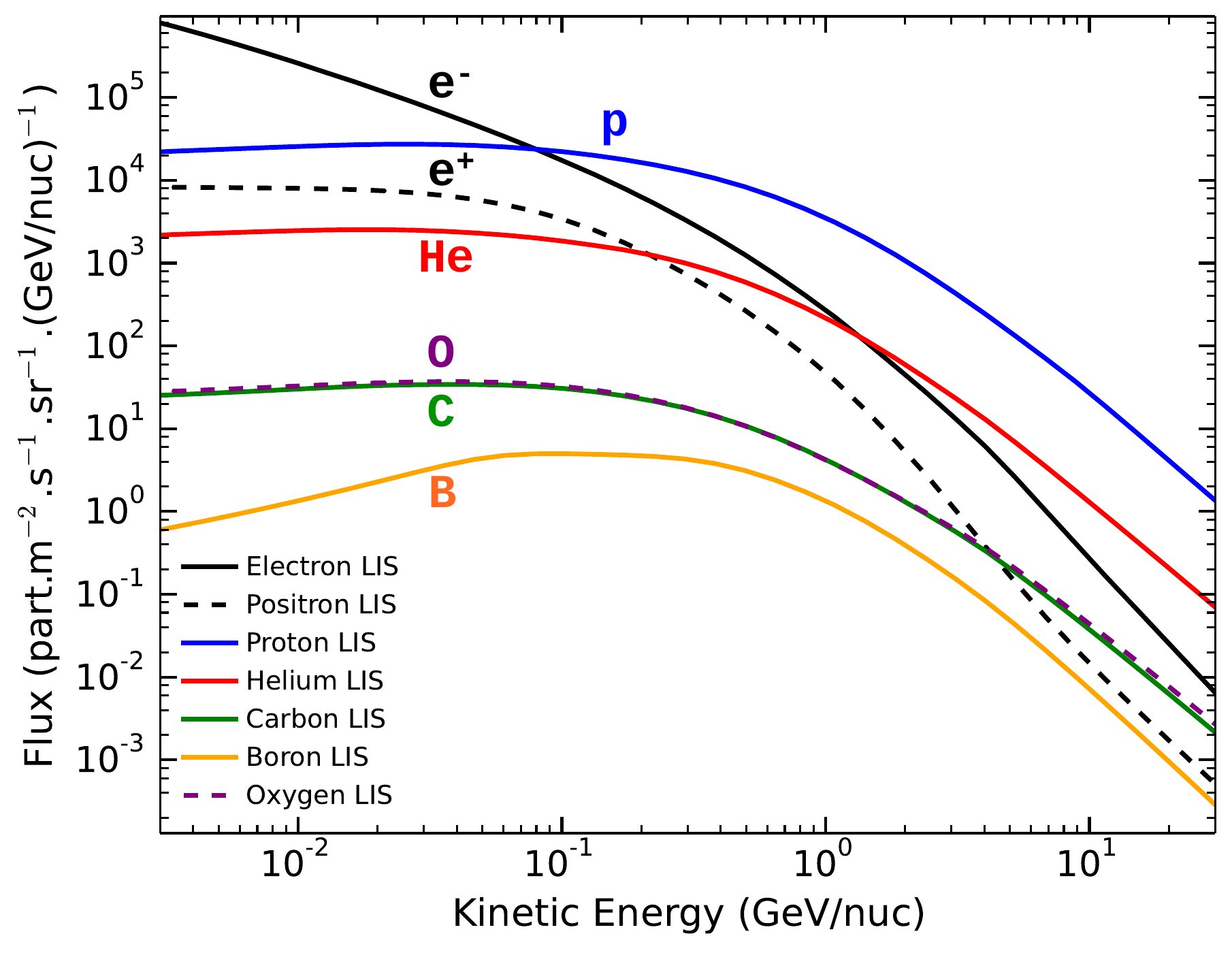}
  \caption{Resulting computed LIS's summarized here based on Fig's. \ref{FinalElec} to \ref{FinalO}. The hadronic CRs are shown as function of kinetic energy (GeV/nuc), together with the electron and positron LIS's on the same axes. Above 0.1 GeV/nuc the protons (solid blue curve) form the bulk of the CRs that arrive at the heliosphere, while below this energy the electrons (solid black curve) take on this role. The computed Carbon LIS (solid green curve) is seen to be nearly identical to that of the computed Oxygen LIS (dashed purple curve).}
  \label{FinalAll}
\end{figure}

We present new LIS's for CR electrons, positrons and protons in Figs. \ref{FinalElec} to \ref{FinalProt} with the corresponding expressions for these LIS's in Eqs. \ref{expr1} to \ref{expr3}.
Similarly the new LIS's for the light CR nuclei Helium, Carbon, Boron and Oxygen are shown in Figs. \ref{FinalHe} to \ref{FinalO}; together with their corresponding expressions Eqs. \ref{expr4} to \ref{expr7}. The corresponding B/C ratio is shown in Fig. \ref{FinalBCrati}. These LIS's were all computed with the GALPROP propagation code using a single model and optimized parameter set as described above. The parameter values are tuned to closely match the respective observations made by V1 and PAMELA, simultaneously for all the above listed CR's. In the energy range at which solar modulation has a significant effect (below about 30 GeV/nuc) the LIS's cannot be directly compared to observations at the Earth. A 3D solar modulation model which takes into account all relevant heliospherical modulation effects, such as particle diffusion, convection and drifts, is therefore used to compute the corresponding CR spectra at the Earth (1\,AU) which are shown with respect to appropriate LIS's and relevant observations as indicated in Figs. \ref{FinalElec} to \ref{FinalO}. In summary, all the new LIS's are shown together in Fig. \ref{FinalAll}. From this comparison the relative differences between the computed LIS's can be seen. Electrons have the largest intensity below about 100\,MeV, while above this energy protons have the largest intensity. This is a consequence of the V1 observations suggesting a power-law for the electron LIS, while the observed proton spectrum levels out instead. The positron LIS has an intensity lower than that of Carbon and Oxygen above about 5\,GeV, but with our current results may exceed that of Helium below 200\,MeV. As expected from the V1 observations, the Carbon LIS is seen to be nearly identical to that of the Oxygen LIS.

In the past the GALPROP plain diffusion model was sufficient when studying electrons \citep{Bisschoff2014}, protons and Helium LIS's separately. Including Carbon, the B/C ratio and positrons into the study, the constraints placed on the LIS's by observations necessitated also considering reacceleration and convection in galactic space. As with the study by  \citet{Bisschoff2016}, the B/C ratio of Fig. \ref{FinalBCrati} more closely matches the ratio observed by PAMELA above 1\,GeV/nuc after taking into account reacceleration. The inclusion of positrons proved the greatest challenge even after also considering the effects of convection, indicating that GALPROP in general might not yet be optimally suited to compute positron LIS's. For future studies antiprotons and the ability of GALPROP to fully compute secondary antiparticles, needs to be investigated in more detail.

Isotopic CR observations from V1 have been studied extensively by \citet{Cummings2016} and such observations at the Earth from ACE by \citet{Lave2013}. Isotopic observations made by PAMELA, such as deuterium, and the extensive range by the AMS-02 experiment could expand these previous studies and what is presented in this work. Heavier CR nuclei can also be considered for follow-up studies, as well as ratios such as p/He, $^3$He$_2$/$^4$He$_2$ and $^{10}$Be$_4$/$^{9}$Be$_4$.

The results presented here are valuable in addition to other galactic and heliospheric studies, such as by \citet{Webber2015elec,Webber2015prot} and by \citet{Herbst2017} who took into account modulation, but used the simplified force-field approximation, which does not take into account important modulation effects such as charge-sign dependence and particle drifts. These force-field models typically over estimate proton spectra below about 1\,GeV during A$<$0 solar magnetic cycles, while underestimating the intensity during A$>$0 cycles. The GALPROP code has also been used by \citet{Cummings2016} to produce LIS's compared to the same observations, but they too only apply a force-field approximation.
Studies have also been presented that attempt to improve on the basic force-field approximation, such as shown by \citet{Corti2016}. A more complicated modulation model has also been presented by \citet{Boschini2017} and similarly to the work presented here use the GALPROP model together with their HELMOD solar modulation model. While these various modeling studies have shown the ability to produce CR LIS's, we believe our 3D solar modulation model to be more refined, and certainly exceeding the force-field approximation. Together with our larger scope of CR LIS's, which also includes positrons, we believe the new LIS's presented here will hopefully form an additional intricate part of future CR and solar modulation studies. It was announced that Voyager 2 crossed the heliopause on November 5, 2018, so that we look forward to see new observational very LIS's from this mission (website: \url{https://voyager.gsfc.nasa.gov/data.html}).

\section*{Acknowledgements}
The authors express their gratitude for the partial funding by the South African National Research Foundation (NRF) under grant 98947. The authors wish to thank the GALPROP developers and their funding bodies for access to and use of the GALPROP WebRun service. D.B. and O.P.M.A. acknowledge the partial financial support from the post-doctoral program of the North-West University, South Africa. This work is dedicated to the memory of William R Webber who passed away at the end of November 2018.

\end{document}